\documentclass[12pt]{article}
\usepackage{epsfig,amsmath}
\begin{document}
\pagestyle{empty}
\baselineskip=0.212in
 
\begin{flushleft}
\large
{SAGA-HE-150-99 
\hfill September 17, 1999}  \\
\end{flushleft}
\vspace{1.0cm}
\begin{center}
\LARGE{{\bf  Proton-deuteron asymmetry}} \\
\vspace{0.2cm}
\LARGE{{\bf  in Drell-Yan processes}} \\
\vspace{0.2cm}
\LARGE{{\bf  and polarized light-antiquark distributions }} \\
\vspace{1.0cm}
\Large
{S. Kumano and M. Miyama $^*$} \\
\vspace{0.2cm}
\Large
{Department of Physics, Saga University, Saga 840, Japan} \\
\vspace{1.5cm}
\Large{ABSTRACT}  
\end{center}

We discuss the relation between the ratio of the 
proton-deuteron (pd) Drell-Yan cross section to
the proton-proton (pp) one
$\Delta_{(T)} \sigma_{pd}/2 \Delta_{(T)} \sigma_{pp}$
and the flavor asymmetry in polarized light-antiquark distributions.
Using a recent formalism of the polarized pd Drell-Yan process,
we show that the difference between the pp and pd cross sections is
valuable for finding not only the flavor asymmetry in longitudinally
polarized antiquark distributions but also the one in transversity
distributions. It is especially important that we point out the
possibility of measuring the flavor asymmetry in the transversity
distributions because it cannot be found in $W$ production
processes and inclusive lepton scattering due to the chiral-odd
property.

\vspace{1.0cm}
\vfill
\noindent
{\rule{6.cm}{0.1mm}} \\
\vspace{-0.2cm}
\normalsize
\noindent
{* Email: kumanos@cc.saga-u.ac.jp, miyama@cc.saga-u.ac.jp.} \\

\vspace{-0.30cm}\noindent
{\ \ \, Information on their research is available at
        http://www-hs.phys.saga-u.ac.jp}  \\

\vspace{0.3cm}
\hfill
{submitted for publication}

\vfill\eject
\pagestyle{plain}

\section{Introduction}

Unpolarized parton distributions are now well known with various
lepton and hadron scattering data. We also have a rough idea
on the longitudinally polarized ones with many experimental
data on the $g_1$ structure function \cite{para}.
However, the details of the polarized distributions are not
known yet. For example, the polarized light-antiquark distributions
are assumed to be flavor symmetric although the flavor asymmetric
sea is confirmed in the unpolarized case \cite{skpr}.
The unpolarized asymmetry was first revealed by the New Muon
Collaboration (NMC) \cite{nmc} in the failure of the Gottfried
sum rule, which was studied in muon deep inelastic scattering
experiments. It was then confirmed by the CERN-NA51 \cite{na51} and
Fermilab-E866 \cite{e866} collaborations in Drell-Yan experiments.
Furthermore, the HERMES semi-inclusive data \cite{hermes}
indicated a similar flavor asymmetry. 
In this way, the $\bar u/\bar d$ asymmetry is now an established 
fact. It is also theoretically understood that various factors
contribute to the asymmetry \cite{skpr}. They include
nonperturbative mechanisms such as meson clouds
and the exclusion principle. Although the effect may not be large,
there could be also a perturbative contribution. 

In order to determine the major mechanism for creating the asymmetry,
other observables should be investigated. The flavor asymmetries
in longitudinally-polarized and transversity distributions
are appropriate candidates for the observables in the light of the
Relativistic Heavy Ion Collider (RHIC) SPIN project and others. 
There are also some model studies on the possible antiquark
flavor asymmetry \cite{model}. Because the $g_1$ data are not enough
to find the asymmetry, we should reply on semi-inclusive or
hadron-scattering ones. For example, charged-hadron production
data are valuable. However, the Spin Muon Collaboration (SMC) and 
HERMES data are not accurate enough at this stage for finding
a small effect although an analysis suggests a slight $\Delta\bar u$
excess over $\Delta \bar d$ \cite{my}. 
There is another possibility of studying it at RHIC by $W$ production
processes. It has been already shown \cite{w-pol,w-unpol}
that the $W$ charge asymmetry is very sensitive to the
antiquark flavor asymmetry in the proton-proton (pp) reaction.
On the other hand, the disadvantage of the $W$ production is that 
the asymmetry in the transversity distributions cannot be
investigated because of the chiral-odd nature \cite{js}.
Therefore, we need to find an alternative method.

This is one of our major purposes for investigating the polarized
proton-deuteron (pd) Drell-Yan processes. An alternative way 
is to combine the pd Drell-Yan data with the pp data as it has been
done in the unpolarized case \cite{na51,e866}. 
However, the formalism of the polarized pd Drell-Yan
had not been available until recently. In particular, it was
not obvious how the additional tensor structure is involved in the
polarized cross sections because the deuteron is a spin-1 hadron.
References \cite{hk1,hk2} made it possible to address ourselves to
the polarized pd processes. Taking advantage of the formalism,
we can discuss the possibility of measuring the polarized
flavor asymmetry by combing the pd Drell-Yan data with 
the pp ones. 
Although such an idea was already pointed out in Refs.
\cite{hk1,hk2}, the purpose of this paper is to show the actual
possibility by numerical analyses.
The relation between the polarized pd Drell-Yan cross
section and the flavor asymmetry is discussed in Sec. \ref{asym}.
Then, numerical results are explained in Sec. \ref{results}, and
conclusions are given in Sec. \ref{concl}.

\section{Flavor asymmetry in polarized proton-deuteron 
         Drell-Yan processes}
\label{asym}

Although the lepton scattering suggests
the asymmetry $\bar u/\bar d \ne 1$ in the failure of
the Gottfried sum rule, it does not enable us to determine
the $x$ dependence. Therefore, the pd Drell-Yan process
($p+d\rightarrow \mu^+\mu^- +X$) has been used for measuring
the unpolarized $\bar u/\bar d$ ratio in combination with
the pp Drell-Yan. Another possibility is to use the $W$ production
processes \cite{w-unpol}. We discuss a method of using the pd Drell-Yan
process for finding the polarized flavor asymmetry in this section.

There are at least two complexities in handing the deuteron reaction.
First, the deuteron is a spin-1 hadron so that additional spin
structure exists. This point is clarified in Refs. \cite{hk1,hk2},
so that the interested reader may read these papers for the details. 
Second, the deuteron structure functions are not
simple summations of proton and neutron ones because of
nuclear effects. Although such nuclear corrections are important
for a precise analysis, we do not address ourselves to them in
this paper because nuclear modification does not affect
major consequences of this paper. 
If experimental data are taken in future, shadowing, D-state admixture,
and Fermi-motion corrections \cite{d-corr} should be taken into account
for detailed comparison.

We found in Ref. \cite{hk1} that the difference between
the longitudinally-polarized pd cross sections is given by
\begin{equation}
\Delta \sigma_{pd}  = \sigma(\uparrow_L , -1_L) 
                       - \sigma(\uparrow_L , +1_L) 
            \propto - \frac{1}{4} \, \left[ 2 \, V_{0,0}^{LL} 
                          + (\frac{1}{3}-cos^2 \theta ) \, 
                            V_{2,0}^{LL} \right]
\ ,
\end{equation}
where the subscripts of $\uparrow_L$, $+1_L$, and $-1_L$
indicate the longitudinal polarization and $\sigma(pol_p,pol_d)$
indicates the cross section with the proton polarization $pol_p$
and the deuteron one $pol_d$. The longitudinally polarized
structure functions $V_{0,0}^{LL}$ and $V_{2,0}^{LL}$ are
defined in Ref. \cite{hk1}.
The subscripts $\ell$ and $m$ of the expression $V_{\ell,m}^{LL}$
indicate that it is obtained by the integration
$\int d\Omega \,  Y_{\ell m} \, \Delta\sigma_{pd}$,
and the superscript $L \, L$ means that the proton and deuteron
are both longitudinally polarized. The $\theta$ is the polar
angle of the lepton $\mu^+$.
A parton model should be used for discussing relations
between the structure functions and parton distributions.
In the following, we take the expression which is obtained
by integrating the cross section
over the virtual-photon transverse momentum $\vec Q_T$.
According to Ref. \cite{hk2}, it is given by
\begin{equation}
\Delta \sigma_{pd} \propto \sum_a e_a^2 \, 
             \left[ \, \Delta q_a(x_1) \, \Delta \bar q_a^{\, d}(x_2)
           + \Delta \bar q_a(x_1) \, \Delta q_a^d(x_2) \, \right]
\ ,
\label{eqn:l-pd}
\end{equation}
where $\Delta q_a^{\, d}$ and $\Delta \bar q_a^d$ are the 
longitudinally-polarized quark and antiquark distributions
in the deuteron. The subscript $a$ indicates quark flavor,
and $e_a$ is the corresponding quark charge. The momentum fractions
are given by $x_1=\sqrt{\tau}e^{+y}$ and $x_2=\sqrt{\tau}e^{-y}$
with $\tau=M_{\mu\mu}^2/s$ and the dimuon rapidity
$y=(1/2)\ln [(E^{\mu\mu}+P_L^{\mu\mu})/(E^{\mu\mu}-P_L^{\mu\mu})]$
in the case of small $P_T$. We neglect the nuclear corrections
in the deuteron and assume isospin symmetry,
so that the distributions in the deuteron become
\begin{alignat}{2}
\Delta u^d = \Delta u + \Delta d, \ \ \ & 
\Delta d^d = \Delta d + \Delta u, \ \ \ & &
\Delta s^d= 2 \Delta s,
\nonumber \\
\Delta \bar u^{\, d} = \Delta \bar u + \Delta \bar d, \ \ \ &
\Delta \bar d^{\, d} = \Delta \bar d + \Delta \bar u, \ \ \ & & 
\Delta \bar s^{\, d} = 2 \Delta \bar s.
\label{eqn:parton-d}
\end{alignat}
The polarized gluon distribution does not contribute to our LO analysis.
Even in the NLO studies, its effect is rather small in the small-$P_T$
region which is investigated in this paper, whereas it contributes
significantly to large-$P_T$ cross sections. Estimates of such gluon
contributions are discussed in Ref. \cite{dy-bgk}.

The situation is slightly different in the transversity case.
If the cross-section difference is simply given by 
$\Delta_T \sigma_{pd} =  \sigma(\phi_p=0,\phi_d=0)
                       - \sigma(\phi_p=0,\phi_d=\pi)$,
where $\phi$ is the azimuthal angle of a polarization vector, 
four structure functions ($V_{0,0}^{TT}$, $V_{2,0}^{TT}$,
$U_{2,2}^{TT}$, and $U_{2,1}^{UT}$) contribute.
Here, the superscript $UT$, for example, indicates that
the proton is unpolarized and the deuteron is transversely
polarized. However, the parton model with the $\vec Q_T$
integration suggests that only $U_{2,2}^{TT}$ remains finite and 
higher-twist functions vanish \cite{hk2}.
In the following discussions, we completely neglect
the higher-twist contributions.
Then, the cross-section difference is given in the parton
model as
\begin{align}
\Delta_T \sigma_{pd} & =  \sigma(\phi_p=0,\phi_d=0)
                      - \sigma(\phi_p=0,\phi_d=\pi)
\nonumber \\
 &  \propto \sum_a e_a^2 \, 
    \left[ \, \Delta_T q_a(x_1) \, \Delta_T \bar q_a^{\, d}(x_2)
          + \Delta_T \bar q_a(x_1) \, \Delta_T q_a^d(x_2) \, \right]
\ ,
\label{eqn:t-pd}
\end{align}
where $\Delta_T q$ and $\Delta_T \bar q$ are quark and antiquark
transversity distributions. The nuclear corrections are again
neglected in the parton distributions of the deuteron, so that
the equations corresponding to Eq. (\ref{eqn:parton-d}) are used
in the following analysis.

The pp cross sections are given in the same way simply by replacing
the parton distributions in Eqs. (\ref{eqn:l-pd}) and (\ref{eqn:t-pd}):
$q^d \rightarrow q$ and $\bar q^{\, d} \rightarrow \bar q$.
The ratio of the pd cross section to the pp one is then given by
\begin{equation}
R_{pd} \equiv \frac{     \Delta_{(T)} \sigma_{pd}}
                   {2 \, \Delta_{(T)} \sigma_{pp}}
        =     \frac{ \sum_a e_a^2 \, 
    \left[ \, \Delta_{(T)} q_a(x_1) \, 
              \Delta_{(T)} \bar q_a^{\, d}(x_2)
            + \Delta_{(T)} \bar q_a(x_1) \, 
              \Delta_{(T)} q_a^d(x_2) \, \right] }
              { 2 \, \sum_a e_a^2 \, 
    \left[ \, \Delta_{(T)} q_a(x_1) \, 
              \Delta_{(T)} \bar q_a(x_2)
            + \Delta_{(T)} \bar q_a(x_1) \, 
              \Delta_{(T)} q_a(x_2) \, \right] }
\ ,
\label{eqn:ratio1}
\end{equation}
where $\Delta_{(T)}=\Delta$ or $\Delta_T$ depending on
the longitudinal or transverse case.
At large $x_F=x_1-x_2$, the $\Delta_{(T)} \bar q_a (x_1)$ terms
can be neglected, so that the ratio becomes
\begin{equation}
R_{pd} (x_F\rightarrow 1) = 1 
- \frac{ [ \, 4 \, \Delta_{(T)} u_v(x_1) \, 
                        -\Delta_{(T)} d_v(x_1) \, ] \, 
   [ \, \Delta_{(T)} \bar u (x_2) - \Delta_{(T)} \bar d (x_2) \, ]}
       { 8 \, \Delta_{(T)} u_v(x_1) \, \Delta_{(T)} \bar u (x_2) \, 
       + 2 \, \Delta_{(T)} d_v(x_1) \, \Delta_{(T)} \bar d (x_2)}
\ ,
\label{eqn:rpd1}
\end{equation}
where $x_1\rightarrow 1$ and $x_2\rightarrow 0$.
If the distribution $\Delta_{(T)}\bar u$ is the same as
$\Delta_{(T)}\bar d$, the ratio is simply given by
\begin{equation}
R_{pd} (x_F\rightarrow 1) = 1 \ \ \
\text{if}\ \Delta_{(T)}\bar u = \Delta_{(T)}\bar d
\ .
\end{equation}
Equation (\ref{eqn:rpd1}) shows that the deviation from one
is directory proportional to
the $\Delta_{(T)} \bar u - \Delta_{(T)} \bar d$ distribution.
If the valence-quark distributions satisfy
$\Delta_{(T)} u_v (x \rightarrow 1) \gg 
 \Delta_{(T)} d_v (x \rightarrow 1)$ \cite{uvdv},
Eq. (\ref{eqn:rpd1}) becomes
\begin{equation}
R_{pd} (x_F\rightarrow 1) = 1 - \left [ \,
    \frac{ \Delta_{(T)} \bar u (x_2) - \Delta_{(T)} \bar d (x_2) }
         { 2 \, \Delta_{(T)} \bar u (x_2) } \, \right ]_{x_2\rightarrow 0}
   =  \frac{1}{2} \, \left [ \, 1 
                 + \frac{\Delta_{(T)} \bar d (x_2)}
                        {\Delta_{(T)} \bar u (x_2)} 
                    \, \right ]_{x_2\rightarrow 0}
\ .
\label{eqn:rpd+1}
\end{equation}
Therefore, if the $\Delta_{(T)} \bar u$ distribution
is negative as suggested by the recent parametrizations and
if the $\Delta_{(T)} \bar u$ distribution is larger (smaller) than
$\Delta_{(T)} \bar d$, the ratio is larger (smaller) than one.
However, if the $\Delta_{(T)} \bar u$ distribution is positive,
it is a different story.
In this way, we find that the data in the large-$x_F$ region
are especially useful in finding the flavor asymmetry ratio
$\Delta_{(T)} \bar u (x) / \Delta_{(T)} \bar d (x)$.

On the other hand, the other $x_F$ regions are not so promising.
For example, if another limit $x_F\rightarrow -1$ is taken,
the ratio is
\begin{equation}
R_{pd} (x_F\rightarrow -1) = 
  \frac{ [ \, 4 \, \Delta_{(T)} \bar u (x_1) \, 
                        +\Delta_{(T)} \bar d (x_1) \, ] \, 
   [ \, \Delta_{(T)} u_v (x_2) + \Delta_{(T)} d_v (x_2) \, ]}
       { 8 \, \Delta_{(T)} \bar u (x_1) \, \Delta_{(T)} u_v (x_2) \, 
        +2 \, \Delta_{(T)} \bar d (x_1) \, \Delta_{(T)} d_v (x_2)}
\ ,
\label{eqn:rpdm1}
\end{equation}
where $x_1\rightarrow 0$ and $x_2\rightarrow 1$.
If the condition $\Delta_{(T)} u_v (x \rightarrow 1) \gg 
 \Delta_{(T)} d_v (x \rightarrow 1)$ is satisfied,
the ratio becomes
\begin{equation}
R_{pd} (x_F\rightarrow -1) = 
     \frac{1}{2} \, \left [ \, 1 
                 + \frac{\Delta_{(T)} \bar d (x_1)}
                   {4 \, \Delta_{(T)} \bar u (x_1)} 
                    \, \right ]_{x_1\rightarrow 0}
\ .
\label{eqn:rpdm2}
\end{equation}
If the antiquark distributions are same, the ratio is
given by $R_{pd}=5/8=0.625$. Comparing the above equation with
Eq. (\ref{eqn:rpd+1}), we find the difference of factor 4.
It suggests that the ratio $R_{pd}$ is not as sensitive as
the one in the large-$x_F$ region although
the $\Delta_{(T)} \bar u/\Delta_{(T)} \bar d$ asymmetry 
could be found also in this region.

The pd/pp ratio has been used for finding the flavor asymmetry
$\bar u/\bar d$ in the unpolarized reaction \cite{na51,e866}.
In this paper, we would like to show the possibility of finding it
in the polarized parton distributions. Our investigation
is particularly important for the transversity distributions.
In finding the flavor asymmetry in the unpolarized and
longitudinally-polarized distributions, popular ideas are to 
use inclusive lepton scattering and $W$ production data.
However, these methods cannot be used for the transversity
distributions because of the chiral-odd property. 
The pd asymmetry in the transversely-polarized Drell-Yan
processes enables us to determine the flavor asymmetry 
$\Delta_T \bar u/\Delta_T \bar d$.

\section{Results}
\label{results}

We show expected pd/pp ratios numerically in this section by
using recent parametrizations for the polarized parton distributions.
First, the leading-order (LO) results are shown in Fig. 1
at $\sqrt{s}=50$ GeV and $M_{\mu\mu}=5$ GeV.
The Drell-Yan cross-section ratio $R_{pd}$ is calculated
in the longitudinally- and transversely-polarized cases,
and the results are shown by the solid and dashed curves,
respectively. The longitudinally-polarized distributions are taken
from the 1999 version of the LSS (Leader-Sidorov-Stamenov)
parametrization \cite{lss}.
Strictly speaking, their distributions cannot be used in the LO
analysis because they are provided at the NLO level.
Nevertheless, the same input distributions are used in our
LO analysis in order to compare with the next-to-leading-order (NLO)
evolution results in the following.
The flavor asymmetry ratio is taken as
\begin{equation}
r_{\bar q} \equiv \frac{\Delta_{(T)} \bar u}{\Delta_{(T)} \bar d}
                = 0.7,\ 1.0,\ {\rm or}\ 1.3,
\end{equation}
at $Q^2$=1 GeV$^2$.
Because the input distributions are provided at $Q^2$=1 GeV$^2$,
they should be evolved to those at $Q^2=M_{\mu\mu}^2$ with
the LO evolution equations \cite{ourcpc} for the longitudinally-polarized
and transversity distributions. 
Although the longitudinal distributions are roughly known from
the $g_1$ data, there is no experimental information on
the transversity ones. Because nonrelativistic quark models
indicate that they are equal to the longitudinal ones,
we assume the same LSS99 distributions at the initial
point $Q^2=1$ GeV$^2$. 
If the antiquark distributions are flavor symmetric ($r_{\bar q}$=1),
the pd/pp ratio satisfies the conditions,
$R_{pd}\rightarrow  1$ as $x_F \rightarrow  1$ and 
$R_{pd}\rightarrow 0.625$ as $x_F \rightarrow -1$.
The flavor-asymmetry effects are conspicuous especially
at large $x_F$ as we explained in Sec. \ref{asym}.
Because the $\Delta_{(T)} \bar u$ distribution is negative
in the used LSS99 parametrization, the ratio $R_{pd}$ is larger
than one at large $x_F$ if there exists a $| \Delta_{(T)} \bar d |$
excess over $| \Delta_{(T)} \bar u |$ ($r_{\bar q}$=0.7).
On the other hand, if $| \Delta_{(T)} \bar u |$ is larger than 
$| \Delta_{(T)} \bar d |$ ($r_{\bar q}$=1.3), it is smaller than one.
In the small-$x_F$ region, the flavor-asymmetry contributions
are not so large due to the suppression factor 1/4.
It is also interesting to find that there is
almost no difference between the longitudinally- and
transversely-polarized ratios if the initial distributions are identical.

\begin{figure}[h]
   \begin{center}
       \epsfig{file=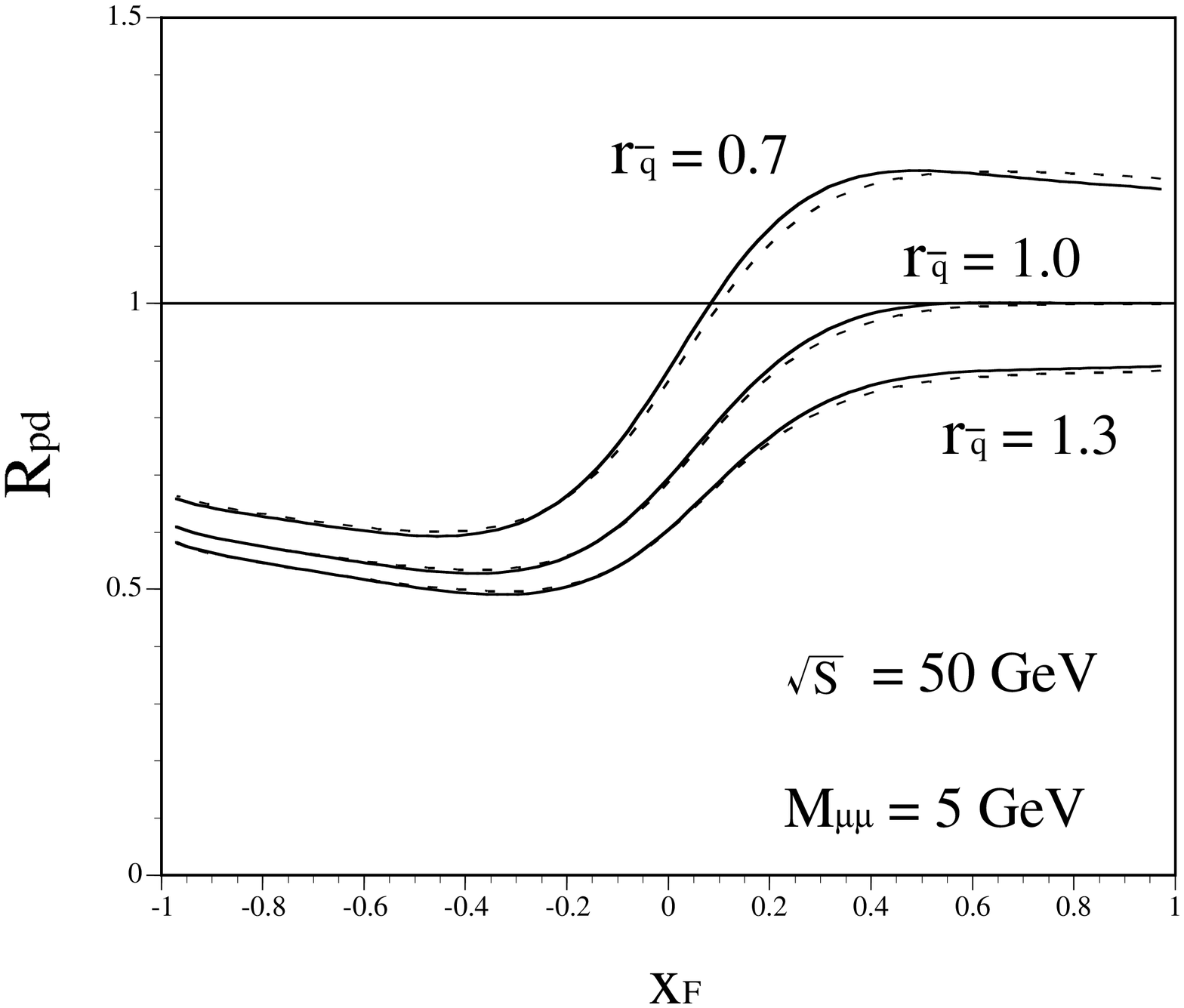,width=9.5cm} 
   \end{center}
   \vspace{-1.21cm}
   \caption{\footnotesize
The Drell-Yan cross section ratio
$R_{pd} \equiv \Delta_{(T)}\sigma_{pd}/2 \Delta_{(T)}\sigma_{pp}$
is calculated in the leading order (LO) of $\alpha_s$ at $\sqrt{s}=50$ GeV
and $M_{\mu\mu}=5$ GeV. The solid (dashed) curves indicate
the longitudinally (transversely) polarized ratios.
The flavor-asymmetry ratio is taken as
$r_{\bar q} \equiv \Delta_{(T)} \bar u / \Delta_{(T)} \bar d$=0.7, 1.0,
or 1.3 at $Q^2$=1 GeV$^2$.
The LSS99 distributions are used for the polarized parton distributions.}
   \label{fig:lo}

\vspace{+0.4cm}
\noindent
   \begin{center}
       \epsfig{file=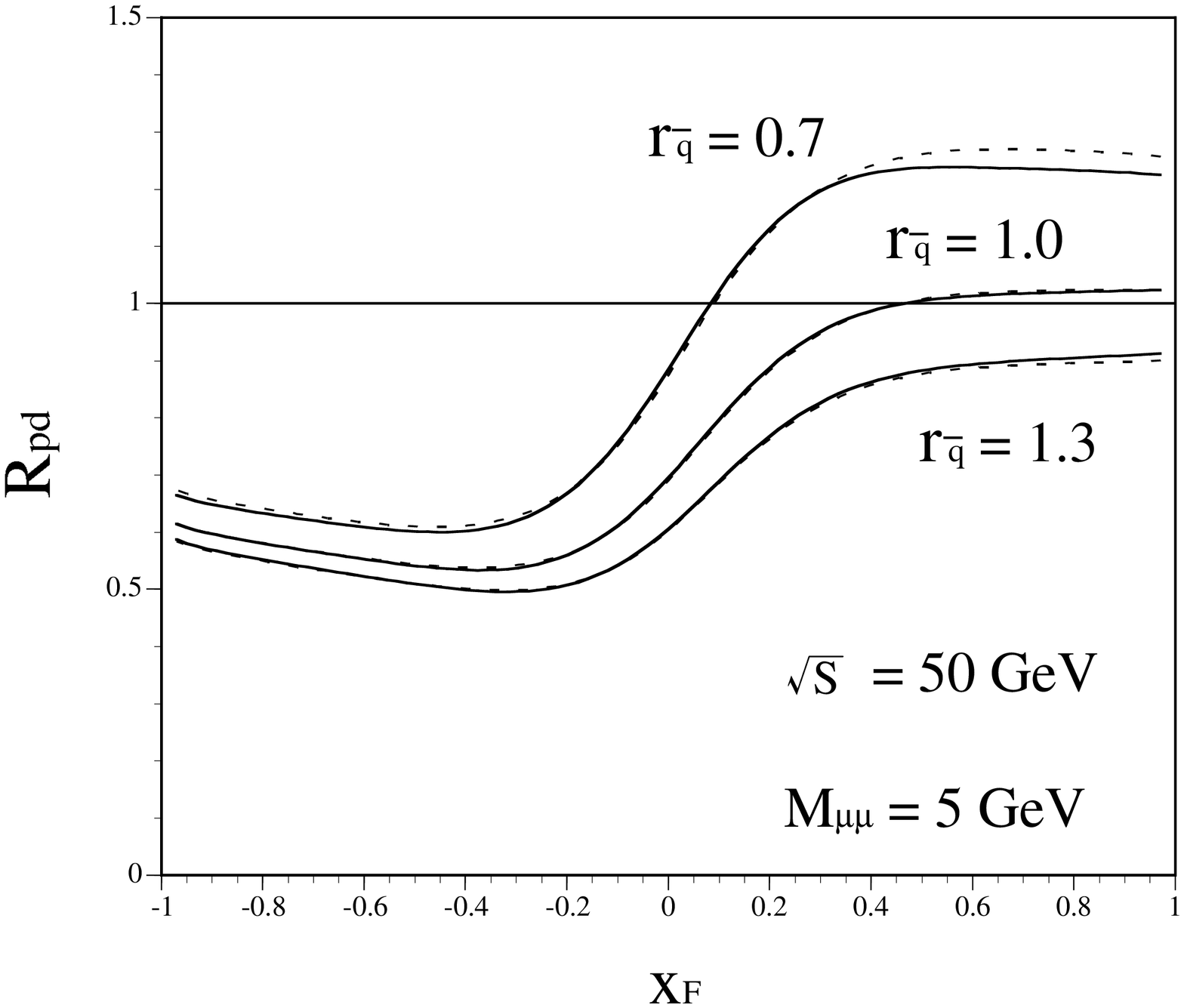,width=9.5cm} 
   \end{center}
   \vspace{-1.21cm}
 \caption{\footnotesize
Next-to-leading-order (NLO) evolution results are shown.
The notations are the same as those in Fig. \ref{fig:lo}.}
 \label{fig:nlo}
\end{figure}

Next, we show NLO evolution results in Fig. \ref{fig:nlo}.
Using the same LSS99 distributions at $Q^2$=1 GeV$^2$, we evolve
them to the distributions at $Q^2$=25 GeV$^2$ by the NLO longitudinal
and transversity evolution equations \cite{ourcpc}.
As shown in the figure, the calculated ratios are almost the same as
those of the LO. However, there are slight differences as it is
noticeable in the large-$x_F$ region: the ratio $R_{pd}$ is not equal
to one although the antiquark distributions are flavor symmetric
at $Q^2$=1 GeV$^2$. It is because the $Q^2$ evolution gives rise
to the asymmetric sea \cite{skpr,ourcpc}
although the initial distributions are flavor symmetric.
This kind of perturbative QCD effect is
not so large in the evolution from $Q^2$=1 GeV$^2$ to 25 GeV$^2$.
If the distributions are evolved from the 
GRSV (Gl\"uck-Reya-Stratmann-Vogelsang) \cite{grsv} type small $Q^2$,
the effect is larger. 
The comparison of Fig. \ref{fig:nlo} with Fig. \ref{fig:lo} indicates
that the NLO analysis is important for a precise determination
of the $\Delta_{(T)} \bar u/\Delta_{(T)} \bar d$ ratio from
measured experimental data.

\begin{figure}[h]
   \begin{center}
       \epsfig{file=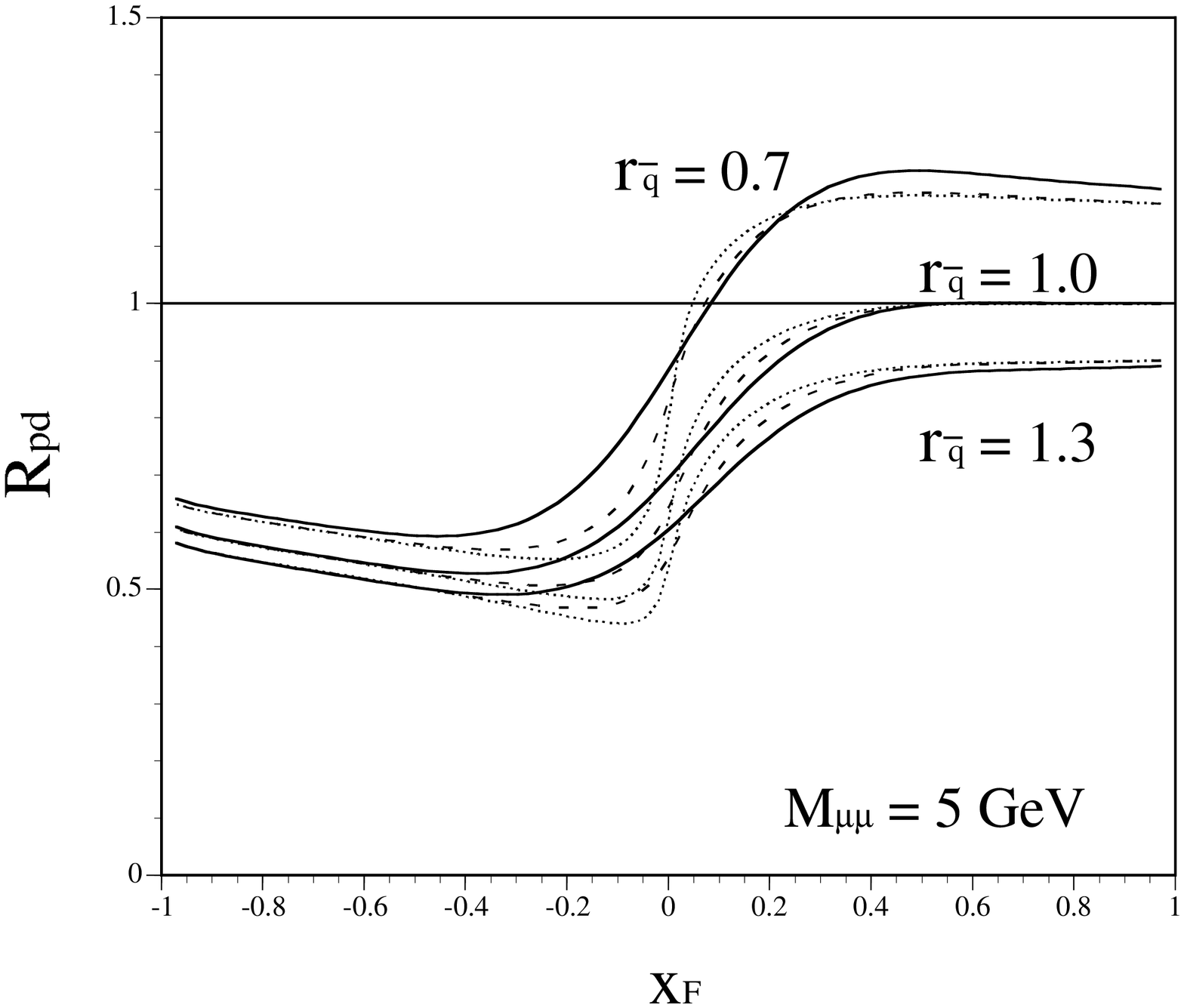,width=9.5cm} 
   \end{center}
   \vspace{-1.21cm}
 \caption{\footnotesize
The dependence on the c.m. energy is shown in the LO. The input
distributions are those of the LSS99.
The solid, dashed, and dotted curves are the longitudinal ratios
at $\sqrt{s}=$50, 200, and 500 GeV, respectively.}
    \label{fig:s}

\vspace{+0.4cm}
\noindent
   \begin{center}
       \epsfig{file=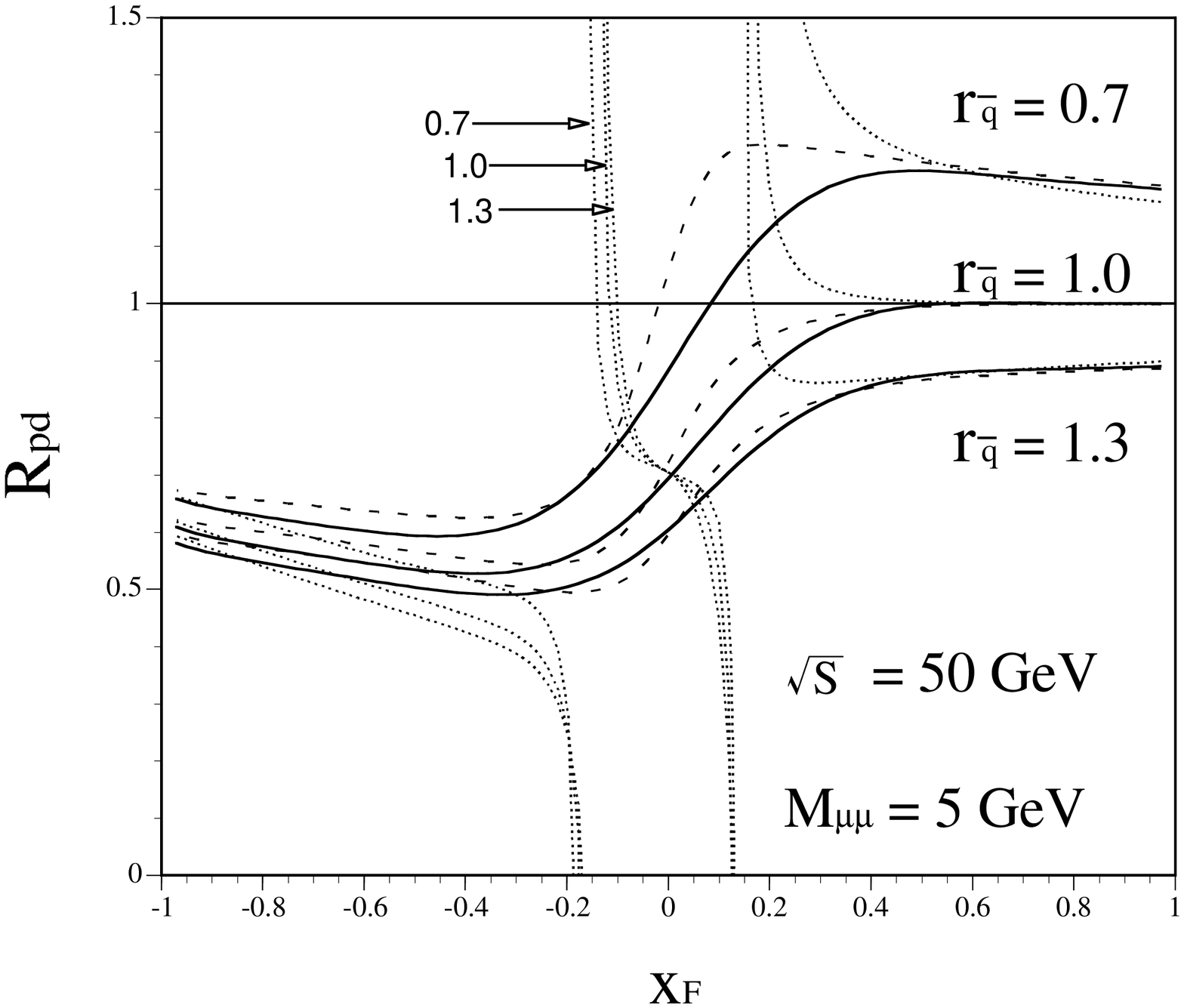,width=9.5cm} 
   \end{center}
   \vspace{-1.21cm}
   \caption{\footnotesize
The dependence on the parametrization is shown in the LO case.
The solid, dashed, and dotted curves are the longitudinal ratios
with the LSS99, GRSV96, and GS-A parametrizations, respectively.}
   \label{fig:para}
\end{figure}

In Fig. \ref{fig:s}, the dependence on the center-of-mass energy
$\sqrt{s}$ is shown. The LO cross-section ratio is calculated at
the RHIC energies $\sqrt{s}$=200 and 500 GeV. 
The calculated ratios are almost equal to those at $\sqrt{s}$=50 GeV
in the large and small $x_F$ regions. 
However, the ratio becomes a steeper function of $x_F$
as $\sqrt{s}$ increases, so that intermediate-$x_F$
results depend much on the c.m. energy. In other words, 
the intermediate region is sensitive to the details of 
the parton distributions.

Finally, we discuss parametrization dependence.
A difference from the unpolarized ratio is that the distributions
$\Delta_{(T)} q$ and $\Delta_{(T)} \bar q$
could be negative so that the denominator,
for example in Eq. (\ref{eqn:ratio1}), may vanish depending
on the kinematical condition. If this is the case, the ratio
has strong $x_F$ dependence in the intermediate region.
Therefore, we should be careful that the obtained numerical results
could change significantly depending on the choice of the input
polarized parton distributions. In particular, the $x$ dependence
of the antiquark distributions, needless to say for the gluon distribution,
is not well known although $\Delta \bar q$ seems to be negative and 
$|\Delta \bar q|$ is rather small according to the recent parametrizations
\cite{para,lss,grsv,gs}.

The LSS99 distributions have been used so far in our analysis.
There are several other polarized parametrizations.
In order to show the dependence on the used parametrization,
we employ the Gehrmann-Stirling set A (GS-A NLO) \cite{gs}
and the GRSV96 (NLO) \cite{grsv}.
The calculated LO ratios are shown in Fig. \ref{fig:para}.
The GS-A and GRSV96 distributions are calculated
first at $Q^2$=1 GeV$^2$ by their own programs.
Then, a certain antiquark ratio $r_{\bar q}$ is introduced. 
After this prescription, the distributions are evolved to
$Q^2$=25 GeV$^2$ by the programs in Ref. \cite{ourcpc}. 
The calculated results are not much different between
the LSS99 and GRSV parametrizations.
However, if the GS-A distributions are used,
the results are much different. This is because the GS-A
antiquark distributions are positive at large $x$ and become
negative at small $x$. The denominator of Eq. (\ref{eqn:ratio1})
could vanish at some $x_F$ points, so that the ratio is infinite at
these points. Therefore, the intermediate-$x_F$ region
is especially useful for finding the detailed $x$ dependence
of the antiquark distributions. 

As we have found in these analyses, the pd Drell-Yan is important
for finding not only new structure functions \cite{hk1,hk2}
but also the details of polarized antiquark distributions.
At this stage, there is no experimental proposal for the polarized
deuteron Drell-Yan. However, there are possibilities at FNAL,
HERA, and RHIC, and we do hope that the feasibility is studied
seriously at these facilities.

\section{Conclusions}
\label{concl}

We have studied the polarized Drell-Yan cross-section ratio
$R_{pd} = \Delta_{(T)} \sigma_{pd} / 2 \, \Delta_{(T)} \sigma_{pp}$.
Using the recent formalism for the polarized
pd processes and typical parametrizations, we have shown that it is
possible to extract the information on the light antiquark flavor asymmetry
($\Delta_{(T)} \bar u / \Delta_{(T)} \bar d$). The large-$x_F$ region
is very sensitive to the flavor asymmetry. Our proposal is particularly
important for the transversity distributions because
the $\Delta_T \bar u / \Delta_T \bar d$ asymmetry cannot be found
in the W production processes. 
Furthermore, the intermediate-$x_F$ region is valuable for finding
the detailed $x$ dependence of the polarized antiquark distributions.

\section*{{\bf Acknowledgments}}
\addcontentsline{toc}{section}{\protect\numberline{\S}{Acknowledgments}}

S.K. and M.M. were partly supported by the Grant-in-Aid for Scientific
Research from the Japanese Ministry of Education, Science, and Culture.
M.M. was supported by a JSPS Research Fellowship for Young Scientists.



\vfill\eject

\noindent{\Large{\bf Figures}}

\vspace{0.4cm}
\noindent

\vfill\eject

\noindent

\end{document}